\documentclass[aps,twocolumn,showpacs,superscriptaddress]{revtex4-1}
\usepackage{amssymb}
\usepackage{amsmath}
\usepackage{graphicx}
\usepackage{hyperref} 
\usepackage{mathtools}
\usepackage[normalem]{ulem}

\DeclarePairedDelimiterX\braket[2]{\langle}{\rangle}{#1\,\delimsize\vert\,\mathopen{}#2}

\usepackage{multirow}
\usepackage{array}
\newcolumntype{P}[1]{>{\centering\arraybackslash}p{#1}}
\newcolumntype{M}[1]{>{\centering\arraybackslash}m{#1}}
\hypersetup{colorlinks=true,
            linkcolor=blue,
            citecolor=blue,
            urlcolor=orange}
        
\usepackage[resetlabels,labeled]{multibib}
\newcites{App}{App Readings}
\usepackage[dvipsnames]{xcolor}

\begin{document}

\title{
Short-range spin correlations and temperature-driven interlayer \\ exchange crossover 
%and persistent s
in the van der Waals magnet CrCl$_3$
%\comAK{Temperature evolution of magnetic properties and} short-range
%\\  magnetic order
%and  in the easy-plane van der Waals magnet $\rm CrCl_3$
}
\author{T. B. Mazitov}
\affiliation{Center for Photonics and 2D Materials, Moscow Institute of Physics and Technology, Institutsky lane 9, Dolgoprudny,
141700, Moscow region, Russia}
\author{M. I. Panin}
\affiliation{Center for Photonics and 2D Materials, Moscow Institute of Physics and Technology, Institutsky lane 9, Dolgoprudny,
141700, Moscow region, Russia}
\affiliation{Russian Quantum Center, Skolkovo, Moscow, 121205, Russia}
\author{A. S. Pakhomov}
\affiliation{Center for Photonics and 2D Materials, Moscow Institute of Physics and Technology, Institutsky lane 9, Dolgoprudny,
141700, Moscow region, Russia}
\affiliation{Russian Quantum Center, Skolkovo, Moscow, 121205, Russia}
\author{M. V. Bakhmetiev}
\affiliation{Russian Quantum Center, Skolkovo, Moscow, 121205, Russia}
\author{E. O. Chiglintsev}
\affiliation{Center for Photonics and 2D Materials, Moscow Institute of Physics and Technology, Institutsky lane 9, Dolgoprudny,
141700, Moscow region, Russia}
\affiliation{Russian Quantum Center, Skolkovo, Moscow, 121205, Russia}
\author{A. I. Chernov}
\affiliation{Center for Photonics and 2D Materials, Moscow Institute of Physics and Technology, Institutsky lane 9, Dolgoprudny,
141700, Moscow region, Russia}
\affiliation{Russian Quantum Center, Skolkovo, Moscow, 121205, Russia}
\author{A. A. Katanin}
\affiliation{Center for Photonics and 2D Materials, Moscow Institute of Physics and Technology, Institutsky lane 9, Dolgoprudny,
141700, Moscow region, Russia}
\affiliation{M. N. Mikheev Institute of Metal Physics, Kovalevskaya Street 18, 620219
Ekaterinburg, Russia.}

\begin{abstract}
We investigate the evolution of spin correlations and the nature of the multi-stage magnetic transition in the quasi-two-dimensional easy-plane van der Waals magnet $\rm CrCl_3$. By combining  broadband {magnetic} resonance spectroscopy  and DC SQUID magnetometry {on mechanically exfoliated micro-flakes} with non-local dynamical mean-field theory (DFT+DMFT) calculations, we 
analyze both long- and short range magnetic order in CrCl$_3$.  
Experimentally, SQUID and {magnetic resonance} measurements confirm the presence of the crossover to a spin polarized phase with the subsequent transition into an antiferromagnetic ground state upon cooling, but show robust short-range correlations at temperatures far above the magnetic ordering temperatures. 
Theoretically, we show the existence of highly stable local magnetic moments at room temperature, with a giant room temperature lifetime $\tau$ of 130--300 ps due to a wide Mott bandgap. Below room temperature, a rapid growth of the in-plane correlation length $\xi$ signals the formation of strong short range magnetic order consistent with the experimental observations. 
We also obtain a temperature-driven crossover of the interlayer exchange interaction, which changes from positive (ferromagnetic) at high temperatures to negative (antiferromagnetic) in the low-temperature ordered phase. 
%establish a direct link between microscopic exchange interactions and macroscopic magnetic phases. Our calculations predict 
%stable ferromagnetic clusters, which create local internal fields. 
%Upon cooling, the FMR linewidth $\Delta H$ exhibits a prominent critical peak at 17 K, which marks the Berezinskii-Kosterlitz-Thouless (BKT) topological transition to an in-plane spin-polarized phase. 
%Below 14 K, the system undergoes a transition to a three-dimensional . 
%Most remarkably, w
%This crossover is driven by a delicate, temperature-dependent balance between ligand-mediated ferromagnetic superexchange and direct chromium-chromium antiferromagnetic exchange.

\end{abstract}

\maketitle

\section{Introduction}

The discovery of long-range magnetic order in atomically thin van der Waals (vdW) crystals has 
%inaugurated 
opened new directions in condensed matter physics, related to low-dimensional magnetism.
%challenging established theoretical boundaries such as the Mermin-Wagner theorem. 
The transition metal trihalides $\rm CrX_3$ ($\rm  X = Cl, Br, I$) have emerged as van der Waals systems for exploring properties of weakly coupled magnetic layers
%quasi-two-dimensional (2D) magnetism 
\cite{macneill2019atomically, ebrahimian2023control, lee2026spin, dupont2021monolayer, soriano2020magnetic, mishra2024magnetic, xue2019two}. While the heavier counterparts, $\rm CrBr_3$ and $\rm CrI_3$, exhibit strong out-of-plane Ising anisotropy \cite{huang2017layer, si2025contrasting, webster2018strain},
%\cite{klein2019enhancement}, 
$\rm CrCl_3$ is characterized by a weak easy-plane XY anisotropy with spins prefer to lie within the honeycomb layers \cite{bedoyapinto2021intrinsic}. This unique configuration makes $\rm CrCl_3$ an ideal platform for observing a wide array of phenomena, including layer-dependent tunneling magnetoresistance \cite{klein2019enhancement,Cai2019} and antiferromagnetic resonance (AFMR) in the gigahertz frequency range \cite{macneill2019gigahertz}. It was also argued that it is characterized by the presence of gapless Dirac magnons in its excitation spectrum \cite{schneeloch2022gapless}.

The magnetic ordering in $\rm CrCl_3$ is notably complex, occurring through a distinctive two-step phase transition \cite{mcguire2017magnetic}. Upon cooling, the so called spin polarized (SP) phase first establishes at approximately 17 K, which is characterized by the appearance of the magnetization in finite  external magnetic fields, which is non-linear with respect to the applied magnetic field.
%This transition was considered as being related mainly to the in-plane correlations. 
With further cooling, this state transitions into the antiferromagnetic (AFM) ground state at $T_N \approx 14$ K at low external fields. Because the interlayer exchange interaction and magnetic anisotropy are much weaker than the intralayer interaction \cite{kim2019evolution}, $\rm CrCl_3$ is highly susceptible to external conditions such as magnetic field, mechanical strain, electrostatic gating, or subtle structural modifications \cite{ebrahimian2023control}.

While the possibility of strong short range magnetic order (SRO) above magnetic transition temperatures was predicted long time ago \cite{Takahashi,Yoshioka}, low magnetic transition temperatures of CrCl$_3$ provide ideal conditions for studying short-range magnetic order  at temperatures significantly exceeding the global ordering point. Previous studies using inelastic neutron scattering and Raman spectroscopy have revealed magnon features \cite{schneeloch2022gapless} and phonon anomalies \cite{kipczak2026strong} that remain identifiable up to at least $100$~K. {Magnetic diffuse neutron scattering was recently observed at least up to 40 K \cite{morgan2026contrasting}.}

Another critical factor influencing magnetic ground state of CrCl$_3$ is the stacking polymorphism inherent to the $\rm CrX_3$ family. Bulk $\rm CrCl_3$ ideally undergoes a first-order structural transition from a high-temperature monoclinic ($C2/m$) phase to a low-temperature rhombohedral ($R\bar{3}$) phase near 240 K. However, stacking defects appear to be important at much lower temperatures \cite{schneeloch2024role}. Also, the preservation of the monoclinic stacking order was reported even at cryogenic temperatures \cite{kesarwani2025investigation}. %Crucially, the $C2/m$ stacking possesses an interlayer exchange interaction stronger than the $R\bar{3}$ phase, leading to significant variations in the saturation fields and magnetic transition temperatures reported across different studies \cite{klein2019enhancement}. Such stacking defects can cause notable discrepancies in magnetization measurements between field-cooled and zero-field-cooled regimes, likely due to the formation of local domains with distinct local critical temperatures .

{Previously, the DFT \cite{besbes2019microscopic,kvashnin2020relativistic, yadav2024electronic} and DFT+DMFT \cite{craco2021electronic} approaches  were applied to describe electronic and magnetic properties of CrCl$_3$. %in trihalides has already proven itself to be reliable in comparison with the single-particle approach which does not cover local dynamical correlations \cite{craco2021electronic}}.
{It was argued, that the intralayer ferromagnetic order (and ferromagnetic correlations in the paramagnetic phase) are caused by the Cl p-states, while the direct exchange of chromium atoms provides only antiferromagnetic contribution \cite{besbes2019microscopic}}.
{The details of the 
%necessity of the anisotropy contribution using Kitaev and 
Dzyaloshinskii-Moriya anisotropic contributions were obtained
%terms to explain the large topological gap and magnon characteristics, as well as the comparison of anisotropic contributions in different trihalides, has been studied in detail previously 
\cite{kvashnin2020relativistic, yadav2024electronic}}. 

The purpose of the present work is to investigate the relation between the long-range and short-range magnetic ordering in CrCl$_3$, as well as the role of stacking defects in the observed magnetic transitions. 
We present a detailed investigation of the temperature dependence of exchange parameters, the evolution of short-range magnetic correlations, and the nature of the multi-stage magnetic phase transition in $\rm CrCl_3$. By combining high-resolution {magnetic} resonance  spectroscopy and DC SQUID magnetometry with the non-local dynamical mean-field theory (DFT+DMFT) calculations, we resolve the hierarchy of interactions in both the monoclinic and rhombohedral phases.  Our results provide microscopic evidence for short range magnetic order and a temperature-driven crossover of the interlayer exchange sign, further establishing $\rm CrCl_3$ as a unique platform for high-frequency magnonics and tunable 2D spintronics.

The plan of the paper is the following. In Sect. II we describe our experimental and theoretical methods. In Sect. IIIA,B we present the experimental results, and in Sect. IIIC theoretical results. In Sect. IV we discuss obtained results and present conclusions.

\section{Methods}

% \begin{figure*}[t]
% \includegraphics[width=1.0\linewidth]{figures/Fig1_clusters.png}
% \caption{(Color online). \textbf{Highly localized short-range magnetic order and magnetic cluster formation in $\text{CrCl}_3$}. \textbf{(Left)} Temperature evolution of the magnetic correlation length ($\xi$), representing the magnetic cluster radius. The inset shows the temperature dependence of $\xi^{-2}(T)$. \textbf{(Right)} Temperature evolution of the intra- and interplane exchange interactions: nearest-neighbor in-plane $J_1$ (red lines/symbols), next-nearest-neighbor in-plane $J_2$ (grey lines/symbols), and nearest-neighbor interplane $J_{\text{inter}}$ (black lines/symbols). Results for the $R\bar{3}$ phase are represented by diamonds and solid lines, while those for the $C2/m$ phase are denoted by triangles and dashed lines. \textbf{(Middle)} Schematic illustration of the $\text{CrCl}_3$ lattice highlighting the formation of magnetic clusters (colored concentric circles, with radii corresponding to $\xi$) and the in-plane exchange pathways (red and grey bonds corresponding to $J_1$ and $J_2$, respectively).}
% \label{fig:clusters}
% \end{figure*}

\subsection{Experimental}

{The studied $\text{CrCl}_3$ flake was mechanically exfoliated on PDMS and then transferred to a $\text{Si/SiO}_2\text{/Pt (6 nm)}$ substrate using a homemade transfer machine.} Flake thickness was equal to 4 $\mu$m and the lateral size was approximately 100 $\mu$m. The sample was covered with polycarbonate to prevent mechanical damage and chemical degradation \cite{panin2025spin}. {The magnetic resonance} was studied using a {VNA-based} setup with a closed-cycle cryostat (Cryotrade CFSG-400-C) \cite{Panin2025}. The coplanar waveguide signal line was narrowed to 100 $\mu$m close to the microwave ports to obtain better excitation efficiency, as it was described in our previous work \cite{panin2025spin}, which also provided the necessary sensitivity to detect resonance modes in micron-scale flakes. The sample was precisely placed on the CPW using markers under optical microscopy control. A 2.5-$\mu$m-thick Mylar film was placed between the sample and signal line to prevent parasitic currents in the Pt layer. The external DC magnetic field ${\mathbf H}_{}$ was directed parallel to the coplanar waveguide (Fig. \ref{fig:waveguide}{a}). Input microwave power was equal 0 dBm in all {magnetic resonance} experiments. $S_{21}$ parameter was analyzed. Frequency was set in the range from 2 to 12 GHz, and the external field was swept from 0.4 T to 10 mT. 

By orienting the in-plane microwave field ${\bf h}_{\rm ip}$ either parallel or perpendicular to the external magnetic field ${\bf H}_{}$, we were able to selectively excite parity-dependent magnon branches. In the parallel configuration (${\bf h}_{\rm ip}~\parallel~{\bf H}_{}$), both optical and acoustic modes were observed below $T_N$. This geometry is particularly advantageous for detecting the optical mode, which represents the out-of-phase precession of the two magnetic sublattices {(Fig. \ref{fig:waveguide}b)} and serves as a definitive signature of the AFM phase \cite{macneill2019gigahertz}. {The acoustic mode, in contrast, corresponds to the in-phase precession of two antiferromagnetically coupled sublattices {(see also Fig. \ref{fig:waveguide}b)} and was studied in a perpendicular configuration (${\bf h}_{\rm ip} \perp {\bf H}_{}$).}

SQUID measurements were conducted using MPMS 5XL Quantum Design. Temperature dependence of the magnetic moment of the sample was measured during heating after zero-field cooling. Isothermal magnetization loops were recorded at various temperatures in the range of 2-60 K to analyze the field-dependent magnetic response.

\begin{figure}[t]
\includegraphics[width=0.8\linewidth]{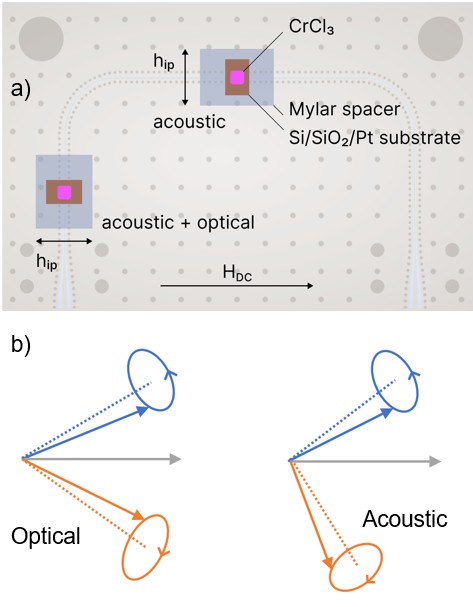}
\caption{a) The scheme of {magnetic resonance} experiment: geometries with ${\bf h}_{\rm ip}$ perpendicular and parallel to the external DC magnetic field ${\bf H}_{}$ are shown; semi-transparent blue rectangles depict Mylar spacer, brown rectangle show Si/SiO$_2$/Pt substrate, and purple circles represents CrCl$_3$ flake under PC coverage, {b) illustration of the in-phase (acoustic mode) and out-of-phase (optical mode) preccession of two antiferromagnetically coupled sublattices, shown with blue and orange arrows.}  }
\label{fig:waveguide}
\end{figure}

\vspace{-0.5cm}
\subsection{Theoretical}
\vspace{-0.25cm}

To obtain theoretical results on the magnetic properties of $\rm CrCl_3$ we use the DFT+DMFT approach. 
For DFT calculations, we use Quantum ESPRESSO \cite{Giannozzi2009, Giannozzi2017} with the standard solid-state pseudopotentials (SSSP PBEsol Efficiency \cite{SSSP}) without considering spin-orbit interaction. We consider the experimentally characterized crystal structures for both monoclinic and rhombohedral structures from \cite{Morosin1964}. The convergence threshold of the DFT self-consistent cycle was $10^{-8}$. The first Brillouin zone was sampled with the $\Gamma$-centered 20×20×20 k-points mesh. The resulting band structures were downfolded to $\rm Cr$ d-orbital states and $\rm Cl$ p-orbital states with Wannier90 package \cite{Pizzi2020}. Next, we studied the electronic many-body correlation effects within the DMFT approach, realized in the Wan2mb\_DMFT software package \cite{Katanin2023,Katanin2025,Wan2mb}, based on the continuous-time hybridization expansion Quantum Monte Carlo (CT-QMC) method for solving the impurity problem \cite{Werner2006}, implemented in the iQIST software package \cite{Huang2015}. Calculation of the non-uniform susceptibilities $\chi_{\mathbf q}$ was performed by the solution of the Bethe-Salpeter equations with an account of 60–80 fermionic Matsubara frequencies, including corrections to finite frequency box   \cite{Katanin2020, Katanin2021} within the Wan2mb\_NL package.

\section{Results}

\subsection{Experimental studies of long-range magnetic order}
%Macroscopic Evidence of Persistent Short-Range Magnetic Ordering}

\begin{figure}[b]
\includegraphics[width=1.0\linewidth]{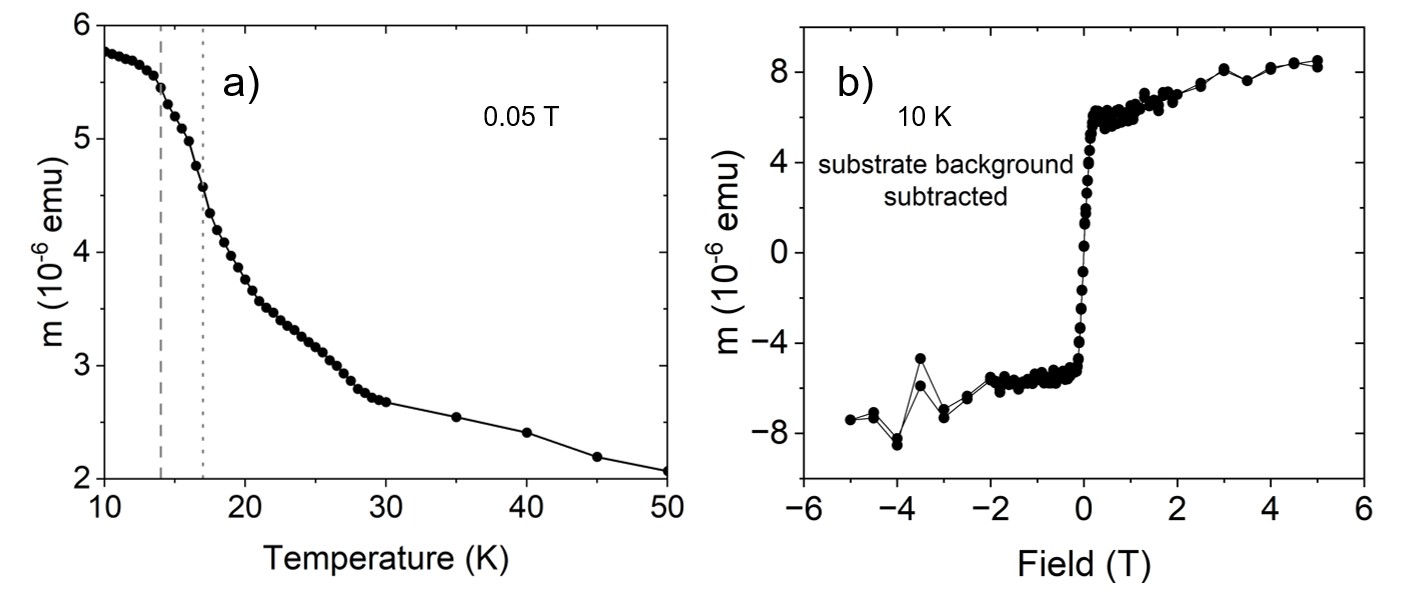}
\caption{SQUID measurements: a) temperature dependence of the sample magnetic moment under 0.05 T in-plane field: dashed line shows the Neel temperature, and dotted line displays the Curie temperature;  b) isothermal hysteresis loop at 10 K. The field is directed in-plane; magnetic field of about 0.2 T is needed to polarize spins.}
\label{fig:SQUID}
\end{figure}

%To further elucidate the nature of the multi-stage magnetic ordering in $\rm CrCl_3$ (and verify some of the previous experimental results), we performed a series of DC magnetic susceptibility and magnetization measurements using a SQUID magnetometer.
{To 
%further 
elucidate the nature of the multi-stage magnetic ordering in $\rm CrCl_3$ (and verify some of the previous experimental results) we conducted DC susceptibility measurements.}
The temperature-dependent magnetization $m(T)$ curves, recorded under an in-plane field of 0.05 T (Fig. \ref{fig:SQUID}a), reveal a sequence of transitions {observed in previous studies \cite{mcguire2017magnetic,kesarwani2025investigation,adma202514405}. In particular,} 
%While the overall signal is affected by the paramagnetic contribution of the substrate, 
a characteristic cusp is clearly visible at $T_N \approx 14$~K, marking the establishment of antiferromagnetic long-range order. This observation is consistent with the established Néel temperature ($T_N$) for bulk $\rm CrCl_3$ \cite{mcguire2017magnetic}. {The inflection point at $T_{\rm SP}\simeq 17$~K is associated with the transition from spin polarized to paramagnetic state}. {We note that despite relatively small magnetic field substantial magnetization is preserved up to $T\sim 50$~K.} {As it is discussed in previous literature \cite{mcguire2017magnetic, schneeloch2024role}, at low temperatures {larger} magnetic field $H\sim {0.2}$T 
%0.1$~T { 0.2 T?} 
induces finite magnetization and the onset of the spin polarized state (see Fig. \ref{fig:SQUID}b)}.

\begin{figure}[t]
\includegraphics[width=1.0\linewidth]{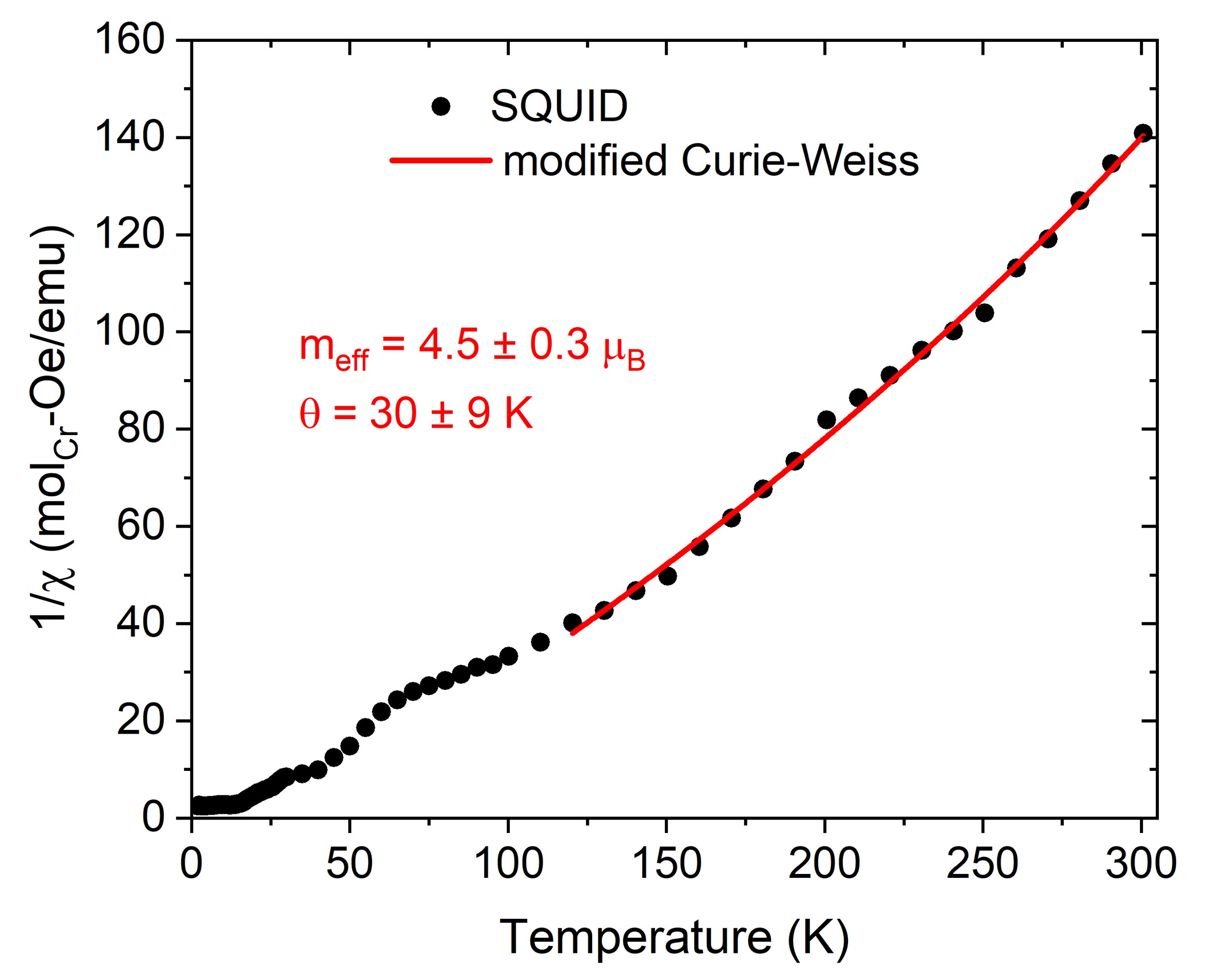}
\caption{Inverse magnetic susceptibility; fit shows modified Curie-Weiss law at high temperatures. }
\label{fig:CW_susceptibility}
\end{figure}

Quantitative analysis of the magnetic state was performed by fitting the inverse susceptibility $1/\chi(T)$ data (Fig. \ref{fig:CW_susceptibility}) to a modified Curie-Weiss law 
\begin{equation}
\chi = \frac{C}{T - \theta} + \chi_0,
\end{equation}
where $\chi_0$ represents the temperature-independent background contribution, $C={N_A \mu_{\rm eff}^2}/({3k_B})$ is the Curie constant, and $\theta$ is the Curie-Weiss temperature. The fit yielded a positive Weiss temperature $\theta \approx 30 \pm 9$ K, which indicates the dominance of ferromagnetic correlations above magnetic transition temperature. 
%exchange interactions within the $\rm CrCl_3$ layers. 
The effective magnetic moment is estimated as $\mu_{\rm eff} \approx \sqrt{8C} \mu_{\rm B} \text{ [cgs]} \approx 4.5 \pm 0.3 \mu_{\rm B}$ {in agreement with previous studies \cite{mcguire2017magnetic} and somewhat higher than the theoretical value for $\rm Cr^{3+}$ ($3.87 \mu_B$).} 
%, a discrepancy that may arise from the non-negligible orbital contribution or the influence of the aforementioned short-range correlations that persist deep into the paramagnetic regime.

\begin{figure*}[t]
\includegraphics[width=1.0\linewidth]{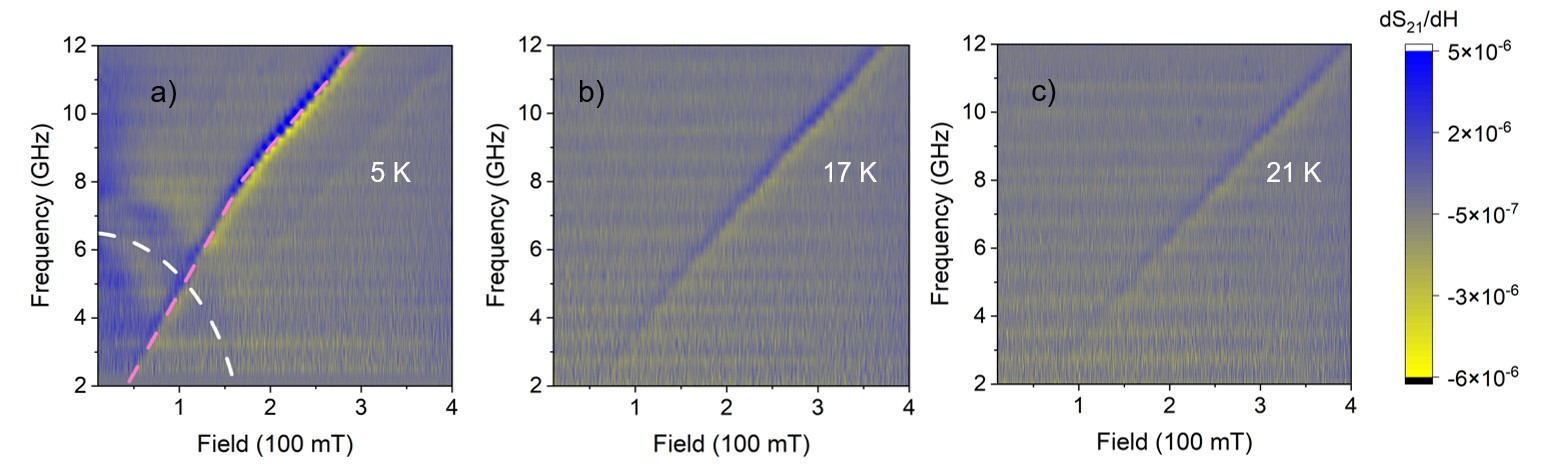}
\caption{a-c) {Magnetic resonance} {intensity} colormaps measured when ${\bf h}_{\rm ip}||{\bf H}_{}$ at 5 K, 17 K and 21 K; acoustic mode is shown with {pink} dashed line and optical mode with {white} dashed line on 5 K colormap.}
\label{fig:FMR_maps}
\end{figure*}

\begin{figure}[b]
\includegraphics[width=1.0\linewidth]{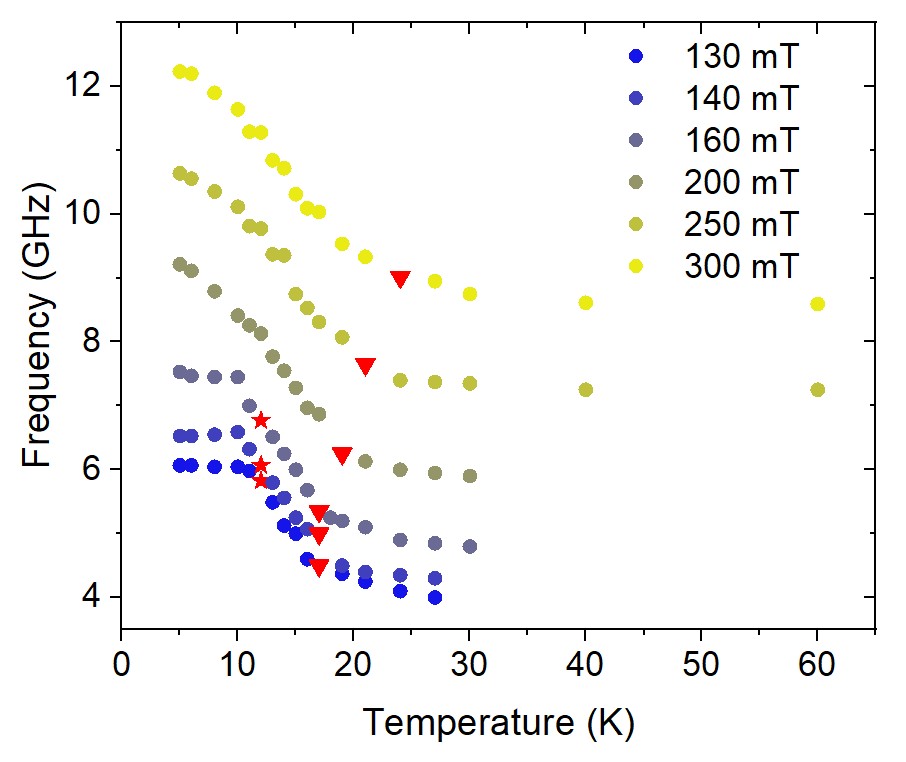}
\caption{Temperature dependence of acoustic and ferromagnetic (Kittel) modes frequencies. Acoustic mode ($H_{}<2H_E)$ frequency is temperature-independent below $T_N$. Stars depict AFM-SP transition and triangles show the SP - SRO-PM crossover. }
\label{fig:FMR_f_T}
\end{figure}

\begin{figure}[t]
\includegraphics[width=1.0\linewidth]{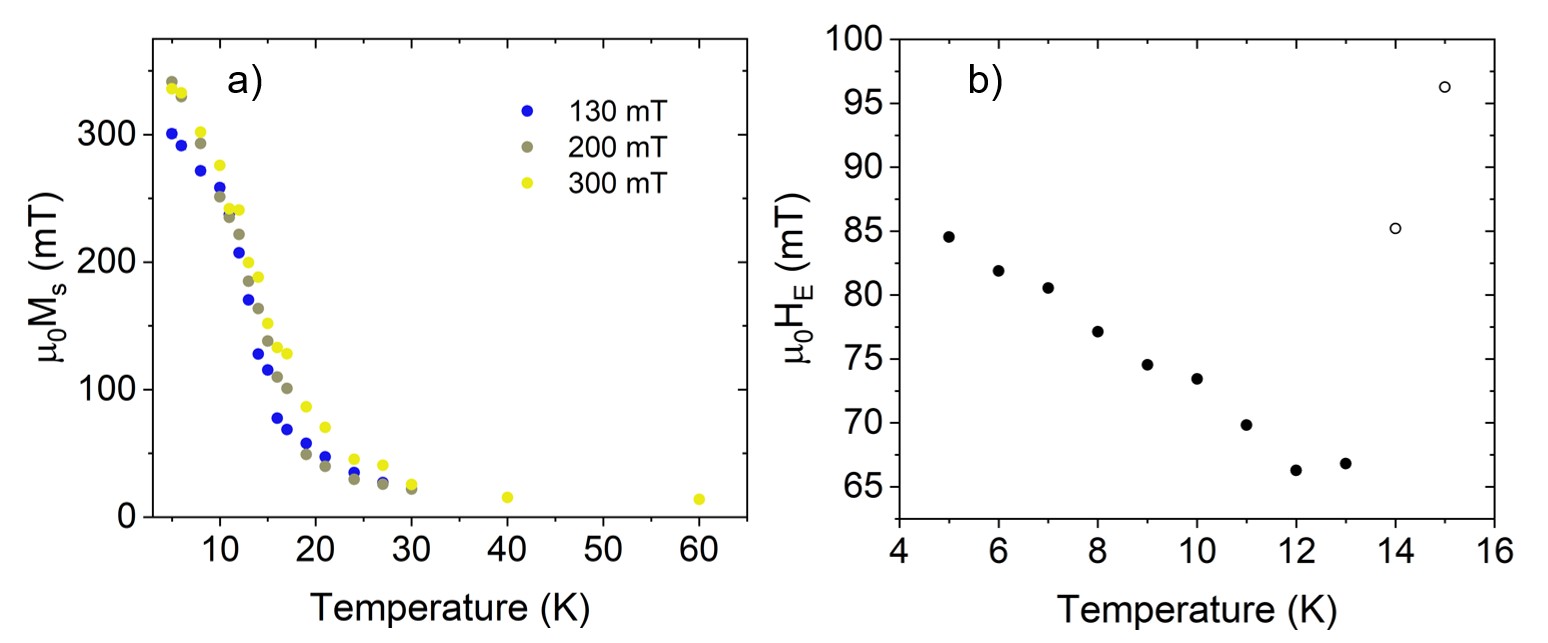}
\caption{Temperature dependence of a) effective magnetization within the layers; b) interlayer exchange field. Open circles in b) correspond to temperatures $T\geq T_N$ (see text).} %{Exchange field shows linear decrease at 5-14 K range, which demonstrates the weakening of the 3D antiferromagnetic order with temperature approaching to 14 K.   We plot 14 K and 15 K $H_E$ data, obtained using the same procedure as the lower temperature data to show the transition, however we consider only $\text{T} < 14 \text{ K}$ values valid. The dependence may also saturate at lower temperatures ( $\sim 2 \text{ K}$), as it was reported earlier \cite{kapoor2021}.} \comTM{Perhaps it would be worth adding vertical lines to indicate the boundaries of the transitions, as in Figure 2?}} 
\label{fig:FMR_temp_dep}
\end{figure}

 To complement the above discussed experimental results 
% and validate our theoretical predictions regarding short-range correlations}, 
we have also performed broadband magnetic resonance measurements in two distinct geometries. 
% The experimental setup utilized a custom coplanar waveguide with a signal line tapered to $100\text{--}\mu\text{m}$ (Fig. \ref{fig:waveguide}), which provided the necessary sensitivity to detect resonance modes in micron-scale flakes. 
 The $\text{S}_{21}$ field derivative colormaps { measured in parallel configuration} (Fig. \ref{fig:FMR_maps}a-c) reveal that below $T_N$, the acoustic {antiferromagnetic resonance (AFMR)} mode eventually transforms into the ferromagnetic Kittel mode at fields exceeding $2H_E$, where $H_E$ is the interlayer exchange field. In the critical range between $T_N$ and $20$~K, the response is dominated by {the ferromagnetic resonance (FMR)} single mode with the non-linear {field response} behavior $f(H)$, while above 20 K, the response becomes purely linear, which is characteristic of the paramagnetic regime, {corresponding to electron paramagnetic resonance (EPR)}.

In the perpendicular configuration (${\bf h}_{\rm ip} \perp {\bf H}_{}$), the excitation of the acoustic mode is significantly enhanced, enabling a high-resolution probe of the transition between the ordered and disordered states up to 50 K (Fig. \ref{fig:FMR_f_T}). We determine the onset of spin polarized phase $T_{\rm SP}$ and antiferromagnetic transition temperatures from the cusps and inflection points of the frequency $f(T)$ dependencies. The AFMR acoustic mode (${H}<2{H}_{\rm E}$) is nearly temperature-independent below $T_N$.  {This stability arises from a compensation of  the exchange field
and shape anisotropy effects \cite{zhang2021zero,sun2025temperature}.} 
To quantify these parameters, we extracted the effective magnetization within the layers $M_s$ (Fig. \ref{fig:FMR_temp_dep}a) and the interlayer exchange field $H_E$ (Fig. \ref{fig:FMR_temp_dep}b) by fitting the frequency-field dependencies. For the canted AFM phase where ${H}_{}<2{H}_{\rm E}$, the acoustic mode data was fitted using:
\begin{equation}
f = \mu_0 \frac{\gamma}{2\pi} \sqrt{2H_E(2H_E + M_s)} \frac{H_{}}{2H_E}.
\end{equation}
{This equation describes the restoring torque governed by the strong interlayer exchange field $H_E$ in conjunction with the layer's magnetic anisotropy},
$M_s$ represents the effective magnetization within the individual layers and $\frac{\gamma}{2\pi} \approx 28 \text{ GHz/T}$ is the gyromagnetic ratio. When the external field exceeds $2H_E$, the sublattices align, and the acoustic mode transforms into the uniform {ferromagnetic} Kittel mode, described by
\begin{equation}
f = \mu_0 \frac{\gamma}{2\pi} \sqrt{H_{}(H_{} + M_s)},
\end{equation}
which is also applicable for SP and SRO-PM phases. 
%$M_s$ is considered to be constant in these equations. 
To obtain the field dependence of the intralayer magnetization, the fit was done with the equations above to obtain $H_E$ and for each field value, $M_s(T)$ was calculated using the obtained $H_E$ for each value of the field $H_{}$. 
The sharp decrease observed in $M_s(T)$ dependencies at 14 K is attributed to AFM-SP transition.  Crucially, $M_s$ remains non-zero at temperatures up to 50 K. This persistent finite magnetization in the absence of global long-range order is a hallmark of strong short-range magnetic order.

{The exchange field $H_E$ decreases with increase of $T$ at sufficiently low temperatures,
%within the $T=5$~-~$14$~K range, 
which accompanies the weakening of the 3D antiferromagnetic order as the temperature approaches Neel temperature $T_N$.
%$T=T_N=14$~K. 
%{
%\sout{We also plot $H_E(T)$ data at $T\gtrsim T_N$ using the same procedure as for the lower temperature data.}} %The respective exchange field shows a sharp increase, which we attribute to the AFM-SP transition; however, we consider only 
%We plot 14 K and 15 K $H_E$ data, obtained using the same procedure as the lower temperature data to show the transition, however we 
%consider only 
%$\text{T} < 14 \text{ K}$ values physically meaningful. 
We 
%considered 
%$T_N$
%the Neel temperature 
%unknown and 
fitted broader temperature range with the procedure valid in the AFM region (Eq. 2,3), obtaining sharp deterioration of the results at %temperatures higher than 
$T>T_N=14$~K (due to the absence of the kink in $f(H)$, shown with open circles in Fig. 6b). This estimate of Neel temperature agrees with that obtained from resonance frequencies in Fig. \ref{fig:FMR_f_T} at low magnetic fields.}
%, which we associate with Neel temperature $T_N$.} 
%connected with the transition to SP phase.}
The temperature dependence $H_E(T)$ may also saturate at lower temperatures ($\sim 2 \text{ K}$), as it was reported earlier \cite{kapoor2021}}.
The resulting values for $H_E$ at low temperatures ($\mu_0 H_E \approx 65\text{--}85$ mT) are characteristic of the rhombohedral ($R\bar{3}$) phase. It was argued \cite{klein2019enhancement} that the monoclinic ($C2/m$) phase possesses an interlayer exchange field nearly ten times larger, confirming that our bulk sample has fully transitioned into the low-temperature rhombohedral structure. 
%corroborating our previous DFT+DMFT calculations that predicted a formal divergence of the correlation length $\xi$ and the formation of stable ferromagnetic clusters well above the global transition point. These clusters create local internal fields that shift the resonance position from the ideal paramagnetic point, effectively preserving "memory" of the FM order in the paramagnetic phase.

\subsection{Persistence of Short-Range Correlations in the Paramagnetic Phase}
\label{Sect:SRO}

%To validate the theoretical prediction of persistent short-range correlations and investigate the multi-stage magnetic phase transition in $\rm CrCl_3$,

{Let us first summarize the evidences of strong short range magnetic order above $T_{\rm SP}$, obtained in previous subsections. 
%From the theoretical side, we have shown persistence of large correlation length and existence of local magnetic moments. Experimentally, t
These findings are related to (i) large magnetic susceptibility above the crossover temperature $T_{\rm SP}$ and (ii) finite in-plane magnetization $M_s$ above $T_{\rm SP}$ in small magnetic fields, obtained by {magnetic resonance}. 

\begin{figure}[t]
\includegraphics[width=0.9\linewidth]{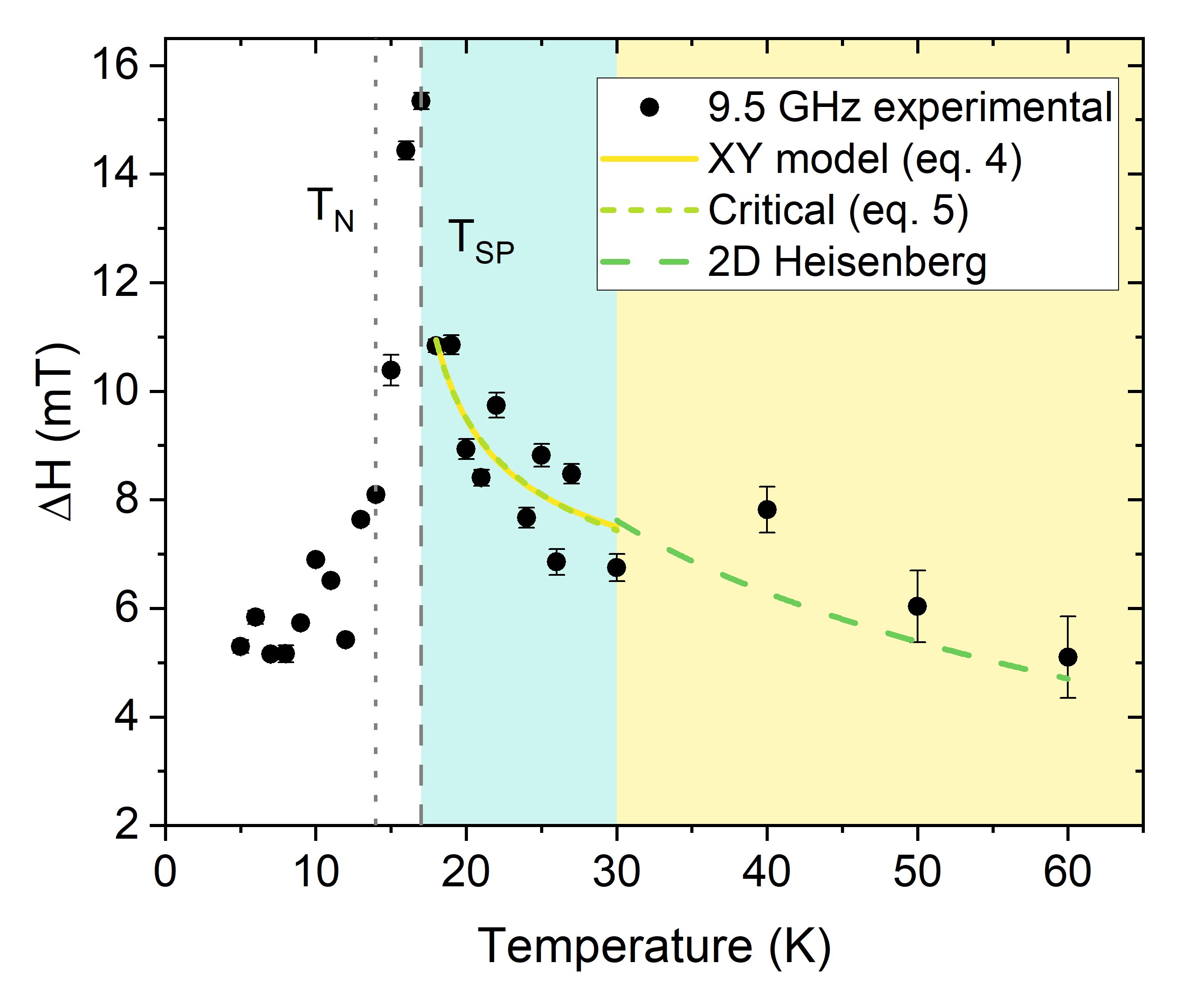}
\caption{Temperature dependence of the resonance linewidth. {{Vertical} dashed and dotted lines show zero-field SRO-PM - SP and SP-AFM transition temperatures respectively. {In blue region {the data are fit by the XY and critical temperature dependence of correlation length}, in yellow region 2D Heisenberg model is more applicable {(see text)}.}}}
\label{fig:FMR_linewidth}
\end{figure}

\begin{figure}[b]
\includegraphics[width=1.0\linewidth]{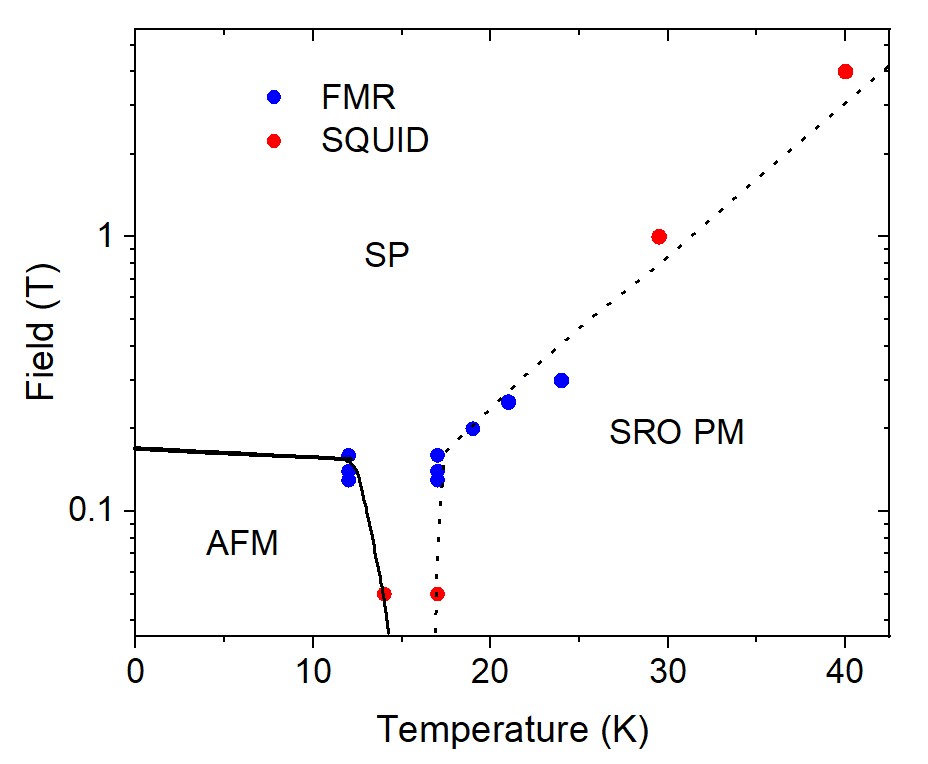}
\caption{$\text{CrCl}_3$ phase diagram {for the in-plane magnetic field} constructed from {magnetic resonance} and SQUID data. Black solid line shows the transition from antiferromagnetic phase (AFM) to spin-polarized phase (SP) and dotted line corresponds to crossover from paramagnetic phase (SRO PM) to spin-polarized.} 
\label{fig:Phase_map}
\end{figure}

The most striking evidence for the evolution of this short-range order is found in the resonance linewidth $\Delta H$ measured at 9.5 GHz (Fig. \ref{fig:FMR_linewidth}). A pronounced peak in the linewidth is clearly observed at $T_{\rm SP}\approx 17$~K. This peak represents a maximum in the spin fluctuation density and signals the transition to the spin-polarized phase. {We note that strong increase of the linewidth with decrease of temperature was also discussed previously \cite{chehab1991two}}, but here it is measured up to the temperature $T_{\rm SP}$. 
%In 2D-XY systems like $\rm CrCl_3$, this temperature corresponds to the Berezinskii-Kosterlitz-Thouless (BKT) transition, where the binding of vortex-antivortex pairs leads to the emergence of quasi-long-range order \cite{bedoyapinto2021intrinsic}.
This increase reflects the rapid growth of magnetic correlation length as the system approaches the transition from the high-temperature side and remarkably remains observable up to at least $60$~K. Slowing down of the spin-dynamics increases the scattering rate of magnons, leading to a broader resonance. The narrowing of the linewidth at low temperatures marks the establishment of coherent long range order, which suppresses the large-scale critical fluctuations, related to AFM long range order. 

{Near the temperature $T_{\rm SP}$ (blue region in Fig. 7, up to 30 K), the behavior of $\Delta H(T)$ can be described by the $\Delta H \sim \xi^3$ relation \cite{Heinrich2003, hemmida2009vortex}, together with the following equation
\begin{equation}
\xi = A \exp \left( b \Big/ \sqrt{\frac{T}{T_{\rm BKT}} - 1} \right)
\label{xsiBKT}
\end{equation} 
for the correlation length $\xi$ of the XY model
and the equation describing the critical behavior near the phase transition,
\begin{equation}
\xi = C\left( \frac{T}{T_{\rm SP}} - 1 \right)^{-\nu}.
\end{equation} 
%(A^3=5.3 mT, b=0.124 , 
Using these equations we obtain 
$T_{\rm BKT}=14.3 K$ and $T_{\rm SP}=16.2 K$.
%(C^3=7.2 mT, p=0.19/3).
{The obtained values of $b=0.12$ and the effective critical exponent $\nu\simeq 0.06$ are anomalously small, which is likely an indication of the presence of structural disorder (discussed also below in Sect. \ref{Exch}) near the transition. This disorder may lead, e.g., to the formation of the Griffiths phase with widening the transition, strongly affecting also the critical behavior \cite{Vojta2003,Pramanik2010}}. % which can originate from structural disorder (see discussion below).}
At higher temperatures (yellow region in Fig. 7), $\Delta H(T)$ is better described using Eqs. (4) and (5) of Ref. \cite{Heinrich2003} for 2D Heisenberg model}. 
%This sequence of transitions: the growth of FM clusters starting at high temperatures, the BKT transition at 18 K, and the final onset of (AFM) long range order at 14 K \--- unifies this experimental observation with the other measurements.

%\subsection{Phase diagram and Temperature-Driven Crossovers of Interlayer Exchange}

By combining the critical temperatures obtained from both SQUID (Fig. \ref{fig:SQUID}a) and {magnetic resonance} (Fig. \ref{fig:FMR_f_T}) measurements, we constructed a comprehensive magnetic phase diagram for $\rm CrCl_3$ under an in-plane field (Fig. \ref{fig:Phase_map}). The diagram reveals three distinct regimes: the 3D AFM phase at the lowest temperatures and fields, the spin-polarized phase, and the short-range ordered paramagnetic phase (SRO PM) at higher temperatures. The boundaries of these phases agree well with existing literature constructed using alternative probes \cite{mcguire2017magnetic,kesarwani2025investigation,Cai2019}. Comparison of the obtained results with Ref. \cite{mcguire2017magnetic} allows us to identify the wide peak of specific heat observed in Ref. \cite{mcguire2017magnetic} {as} the crossover from SRO PM to the SP phase, while the narrow low temperature peak of specific heat corresponds to the magnetic transition to the AFM phase. We also note that at {weak} magnetic fields $H<H_{\rm cr}\simeq 0.2$~T the temperature $T_{\rm SP}$ only {marginally} depends on magnetic field, while for $H>H_{\rm cr}$ we find an increase of $T_{\rm SP}\simeq T_1\ln(H/H_0)$ with {$T_1\simeq{16.3}$}~K and $H_0{\simeq{0.014}}$~T being constants. 

%The unique stability of the SRO PM region, evidenced by both the persistent hysteresis loops and the frequency shifts in FMR, confirms that $\rm CrCl_3$ acts as a prototypical 2D XY system where the magnetic state is governed by highly localized and robust short-range order.

\subsection{Theoretical analysis of magnetic properties}

\begin{figure}[b]
\includegraphics[width=1.0\linewidth]{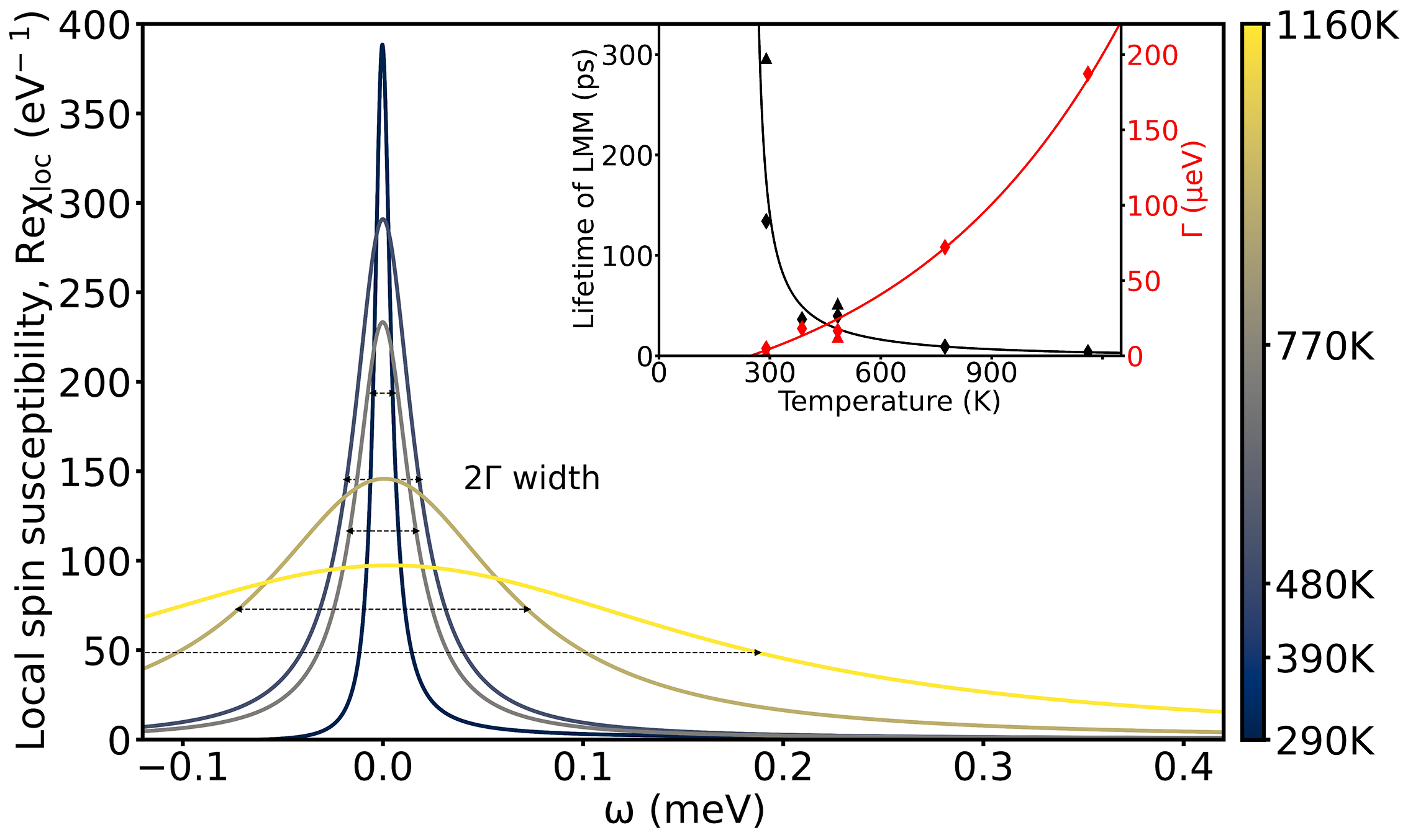}
\caption{(Color online). %Giant lifetime of local magnetic moments (LMMs) in $\text{CrCl}_3$. 
Temperature evolution of the dynamic local spin susceptibility, $\text{Re}\,\chi_{\rm loc}(\omega)$. The line color corresponds to the temperatures indicated on the colorbar on the right. 
%The horizontal dashed lines with arrows illustrate the full width at half maximum ($2\Gamma$). 
The inset shows the temperature dependence of the peak half-width $\Gamma$ (red, right axis) and the resulting {local magnetic moment (LMM)} lifetime, $\tau = \hbar / \Gamma$ (black, left axis). Results for the $R\bar{3}$ phase and the $C2/m$ phase are represented by diamonds and triangles, respectively.}
\label{fig:LMM}
\end{figure}

\subsubsection{Local magnetic moments and correlation length}

{Magnetic properties of CrCl$_3$ can be effectively described within the local magnetic moments picture.} Although in the DFT approach, CrCl$_3$ appears to be metallic, in the DFT+DMFT approach, the gap $\sim 3.5$~eV in the electronic spectrum is opened by correlation effects (mainly by Hund exchange), as discussed in a recent study \cite{ermolaev2025giant}. In spite of this, the temperature evolution of the real part of the dynamic local spin susceptibility, $\text{Re}\,\chi_{\rm loc}(\omega)$, in CrCl$_3$ shows very narrow peaks, whose width strongly decreases with temperature (Fig. \ref{fig:LMM}). This highlights stable local moments below room temperature. Notably, even at the highest temperature $T\sim 1000$~K, the peak half-width $\Gamma$ is only about 0.2 meV, such that $\Gamma \ll kT$ at all considered temperatures, indicating strong localization and weak damping of local magnetic moments.
%, the local magnetic moments (LMMs) remain remarkably well-defined even under extreme thermal agitation, possessing a lifetime $\tau$ of 
The corresponding lifetime of local magnetic moments increases from 3 ps at high temperatures to ~300 ps at room temperature 
%Cooling toward room temperature (290 K, represented by the dark blue curves) suppresses this damping 
increasing by nearly two orders of magnitude.}

\begin{figure}[t]
\includegraphics[width=0.85\linewidth]{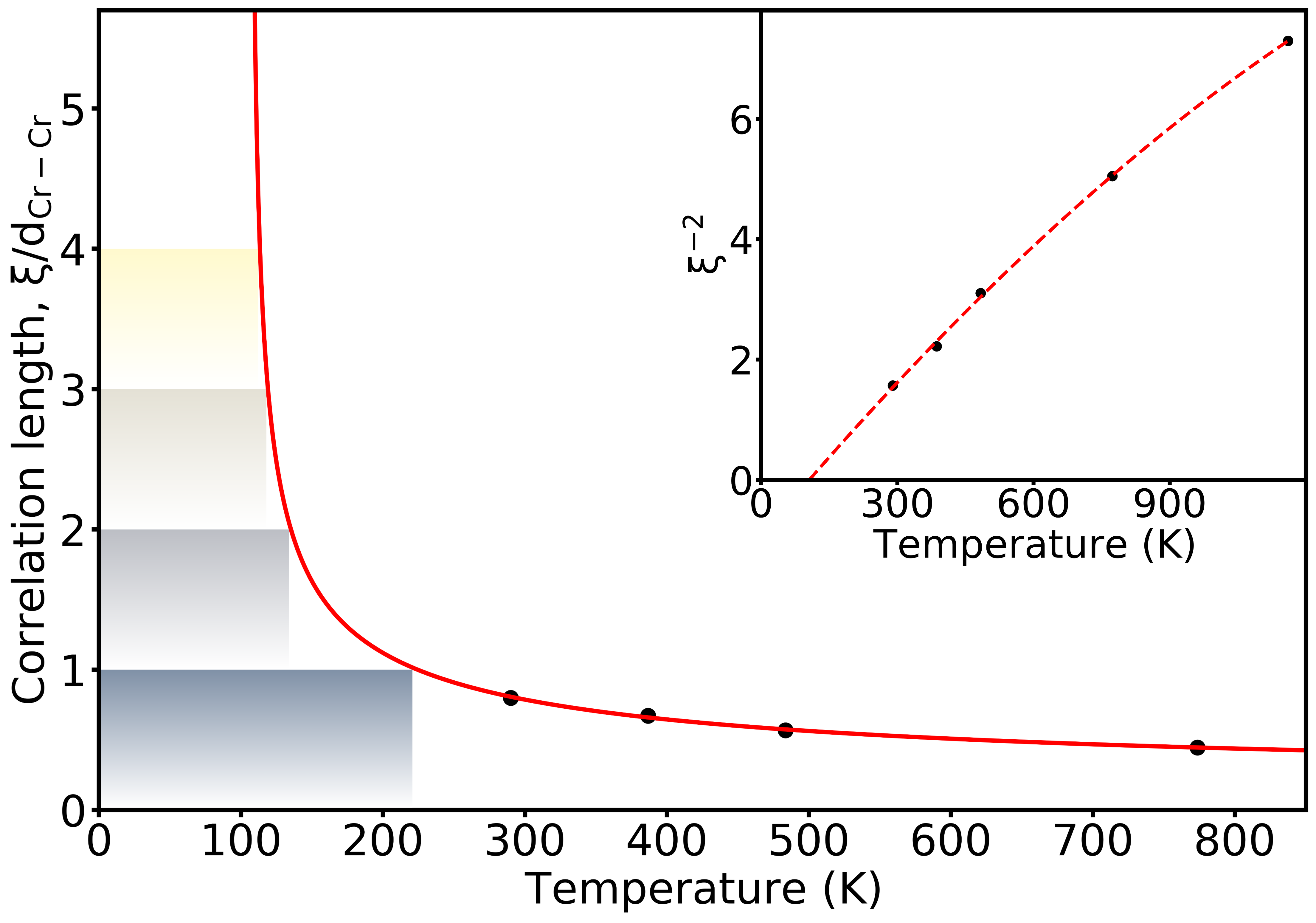}
\caption{(Color online). %Magnetic cluster formation in $\text{CrCl}_3$. 
Temperature evolution of the magnetic correlation length $\xi$ (in units of the nearest distance between chromium atoms $d_{\rm{Cr-Cr}}$).
%, representing the magnetic cluster radius. 
%The colored areas correspond to the sizes of the clusters in the schematic illustration in Fig. \ref{fig:exchange}. 
The inset shows the temperature dependence of $\xi^{-2}(T)$.}
\label{fig:corr_length}
\end{figure}

%The results of our theoretical analysis, performed using DFT+DMFT methods including non-local expansions, provide a comprehensive picture of the magnetic state evolution in $\rm CrCl_3$. 
%We consider first the correlation length. 
{To characterize the spatial correlations of spins at different chromium atoms, in} Fig. \ref{fig:corr_length}, we present the calculated temperature dependence of the in plane magnetic correlation length $\xi$, determined from fitting the spin susceptibility to Ornstein-Zernike form $\chi_{\mathbf q}\propto(q^2+\xi^{-2})^{-1}$. The obtained correlation length serves as a quantitative measure of the linear size of magnetic clusters. At room temperature, $\xi$ remains relatively small, but it demonstrates a rapid increase {at lower} temperature{s} {signaling establishing of magnetic correlations}. The inset of Fig. \ref{fig:corr_length} showing the inverse correlation length $\xi^{-2}(T)$ reveals a formal divergence at $T^* \simeq 100$ K. While this value is significantly higher than the experimental crossover temperature ($T_{\rm SP} \approx 17$~K), 
%such an overestimate is a well-known artifact 
in view of the mean-field nature of DMFT, this divergence 
should rather be interpreted as a temperature scale for the establishment of strong short range magnetic order. {Because of using the DMFT approach and the procedure of extrapolation of the results to low temperatures, this estimate of the temperature $T^*$ remains qualitative, rather than quantitative, but it represents an `order of magnitude' agreement with the 
temperature of establishing short range magnetic order observed in Sect. \ref{Sect:SRO}, as well as it agrees with recent observations of magnetic excitations in $\rm CrCl_3$ that remain visible up to $T\sim 100$~K \cite{schneeloch2022gapless}}.
%tends to overstabilize ordered phases (especially in low-dimensional systems) by neglecting long-wavelength transverse fluctuations. The obtained temperature scale represents the onset of robust short-range 
%ferromagnetic correlations within the layers. 
We also note that, since the excitons in CrCl$_3$ are well localized, even magnetic correlations with $\xi\gtrsim d_{\rm Cr-Cr}$ may affect optical properties
\cite{ermolaev2025giant}. 
%The persistence of these correlations well into the paramagnetic phase is consistent with the  

\begin{figure}[b]
\includegraphics[width=1.0\linewidth]{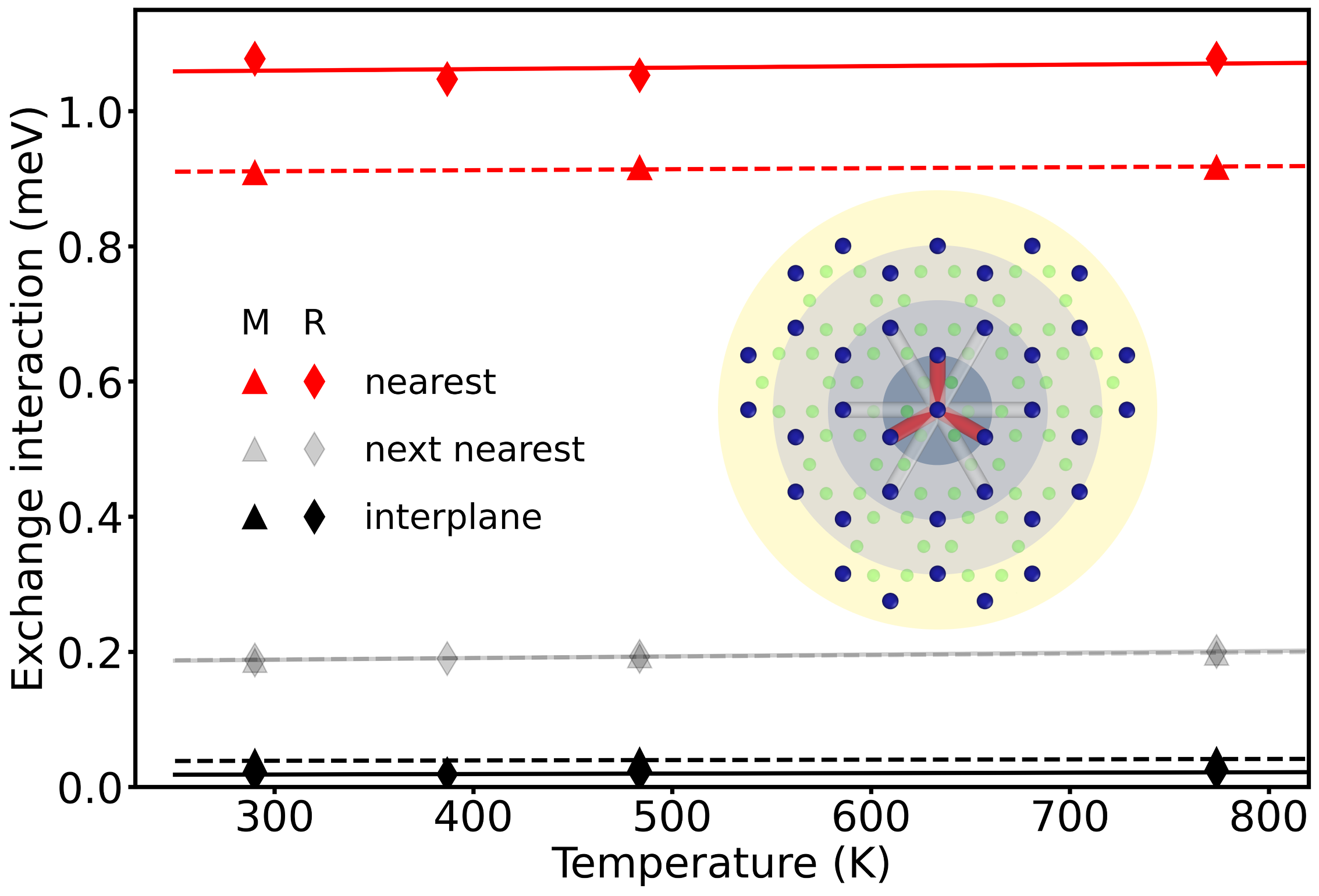}
\caption{(Color online). %Highly localized short-range magnetic order in $\text{CrCl}_3$. 
Temperature evolution of the intra- and interplane exchange interactions: nearest-neighbor in-plane exchange $J_1$ (red lines/symbols), next-nearest-neighbor in-plane exchange $J_2$ (grey lines/symbols), and nearest-neighbor interplane exchange $J_{\text{inter}}$ (black lines/symbols). Results for the $R\bar{3}$ phase are represented by diamonds and solid lines, while those for the $C2/m$ phase are denoted by triangles and dashed lines. Inset shows coordination spheres
%$\text{CrCl}_3$ lattice highlighting the formation of magnetic clusters 
(concentric circles are marked by the color, corresponding to various $\xi$ values from Fig. \ref{fig:corr_length}); 
%and the in-plane exchange pathways (
red and grey bonds correspond to nearest- and next nearest neighbors, respectively.}
\label{fig:exchange}
\end{figure}

%\subsubsection{In plane magnetic exchange and BKT temperature}

\subsubsection{Exchange interactions and BKT temperature}
\label{Exch}

{To characterize magnetic interactions between chromium atoms, we have calculated exchange parameters $J_{ij}$ of the effective Heisenberg model as $J_{\mathbf q}=
\chi_{\rm loc}^{-1}(0)-\chi_{\bf q}^{-1}$ \cite{Katanin2023,Katanin2023Co},
 where the matrix inverse is taken with respect to the site indexes in the unit cell. The temperature dependence of exchange interactions} (Fig. \ref{fig:exchange})  elucidates the stability of local magnetic moments. The intralayer exchange with the  nearest neighbor ($J_1$) is approximately $0.9\text{--}1.1$ meV, which is nearly an order of magnitude stronger than the interaction with the next nearest neighbor ($J_2$). This sharp decay indicates a high degree of magnetic moment localization on the $\rm Cr^{3+}$ ions. Furthermore, these exchange interactions remain nearly constant across a wide temperature range, confirming the applicability of the effective Heisenberg model.

A comparison between the structural phases reveals that the rhombohedral $R\bar{3}$ phase, which is the stable low-temperature modification, possesses higher intralayer exchange values than the high-temperature monoclinic $C2/m$ phase. This suggests that the structural phase transition at approximately 240 K provides an additional energetic "boost" to the magnetic subsystem, enhancing the stability of short range magnetic order. Conversely, the interlayer exchange ($J_{\perp}\simeq 0.02$-$0.04$~meV) is found to be 25-50 times weaker than $J_1$. This 
%quantitative difference 
underscores the quasi-2D nature of $\rm CrCl_3$ and explains the dominance of 2D short range order at not too low temperatures.
%\cite{klein2019enhancement}.

Since $\rm CrCl_3$ exhibits easy-plane magnetic anisotropy \cite{si2025contrasting, webster2018strain, bedoyapinto2021intrinsic}, we apply the estimate for the Berezinskii-Kosterlitz-Thouless transition temperature \cite{irkhin1999kosterlitz}
% CrCl$_3$ possesses the easy-plane anisotropy  To calculate Berezinskii-Kosterlitz-Thouless (BKT) transition temperature we use the formula 
\begin{equation}
T_{\rm BKT} = \frac{4 \pi \rho_s}{\rm ln (T_{\rm BKT}/|\Delta|) + 4 \rm ln (4 \pi \rho_s / T_{\rm BKT})},
\end{equation}
where $\Delta(T) = (2S-1)(E_{\rm sia} + E_{\rm dip})$, 
% $\rm MAE$ is the magnetic anisotropy energy 
where $E_{\rm sia}$ and $E_{\rm dip}$ are the single-ion and dipolar anisotropy parameters, $\rho_s = D S/A$ is the spin stiffness, $D$ is the spin-wave stiffness, $A$ is the cell area. For $E_{\rm sia} + E_{\rm dip} = 
 -31.4~\mu{\rm eV}+ 85.2~\mu{\rm eV} = 
53.8~\mu{\rm eV}$ \cite{yadav2024electronic}, {$D = 17$~meV$\cdot \AA^2$ }, obtained from the calculated exchange interactions, and spin $S = 1.5$ ($\rm Cr^{3+}$) we obtain a value of {$T_{\rm BKT} = 18$~K}  for the $R\bar{3}$ phase, which is in a reasonable agreement with the experimental data.
%and $T_{BKT} = 16 K$ for the $C2/m$ phase. 
%Given that $\rm CrCl_3$ is characterized by easy-plane XY anisotropy, our results suggest that the transition at 17 K marks the topological binding of vortex-antivortex pairs within the individual layers. 
%The respective magnetic transition temperature is expected to be slightly larger and determined by the interlayer exchange interaction \cite{irkhin1999kosterlitz}. %This confirms that the observed magnetism in $\rm CrCl_3$ is fundamentally a 2D phenomenon, while the transition to a 3D 
%The low-temperature antiferromagnetic state is considered in the following.
%at slightly lower temperatures is a secondary effect of weak interlayer coupling, which we will discuss further.

%The predicted stability of these localized ferromagnetic clusters well above the global transition point sets the stage for their experimental detection via high-frequency magnetic resonance.

\begin{figure}[b]
\includegraphics[width=1.0\linewidth]{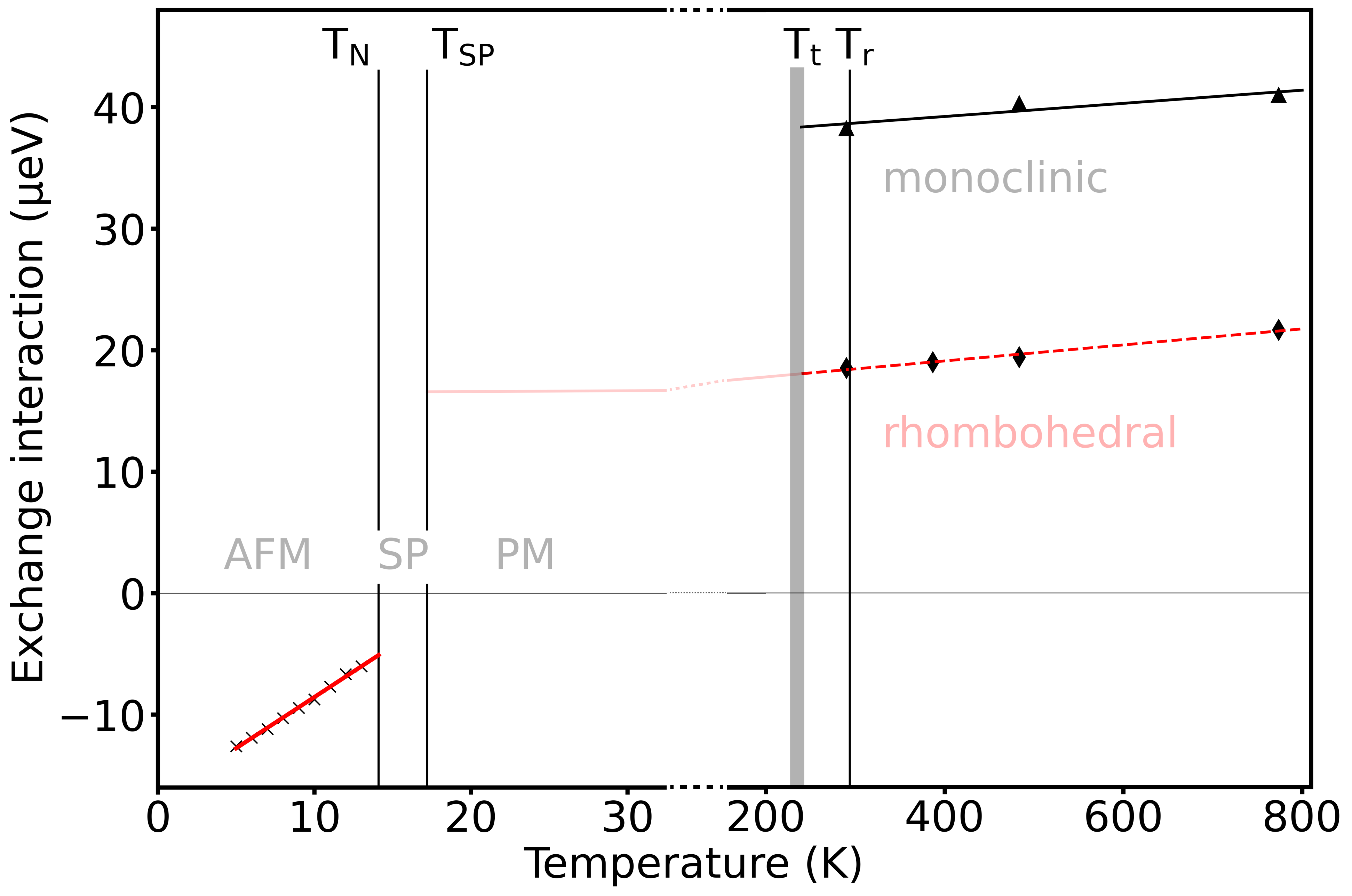}
\caption{(Color online). Interlayer exchange interaction with nearest interlayer neighbors in $\rm CrCl_3$ in DMFT approach (black triangles and rhombuses) and interlayer exchange interaction ($H_E$) in {magnetic resonance} experiment (black crosses). The room temperature ($T_{r}=294 K$), the temperature of the structural phase transition ($T_t=230-240 K$), the temperature of the magnetic transition to the spin-polarized phase ($T_{\rm SP}=17 K$), and the temperature of the magnetic transition to the AFM phase ($T_N=14.1 K$) are also marked.}
\label{fig:interlayer_exchange}
\end{figure}

A comprehensive investigation of the exchange interactions between the layers, summarized in Fig. \ref{fig:interlayer_exchange}, provides insight into the competition between ferromagnetic and antiferromagnetic tendencies in $\rm CrCl_3$. Our DFT+DMFT calculations in the paramagnetic (PM) phase demonstrate that at high temperatures, the interlayer exchange is positive (FM). As the system cools toward room temperature, the magnitude of this FM interaction is gradually suppressed for both the monoclinic and rhombohedral phases, indicating a weakening of high-temperature ferromagnetic fluctuations. 

While the DMFT calculations are primarily targeted at the paramagnetic regime, extrapolating the exchange behavior toward lower temperatures, we expect a significant jump upon the transition from the high-temperature monoclinic ($C2/m$) to the low-temperature rhombohedral ($R\bar{3}$) symmetry. This jump underscores the extreme sensitivity of the interlayer coupling to the stacking registry. Such structural control over magnetism has been previously highlighted in exfoliated flakes, where the inhibition of the $M \to R$ transition can lead to a nearly tenfold enhancement of the exchange interaction compared to the bulk rhombohedral state \cite{klein2019enhancement}.

Remarkably, the strength of the obtained interlayer exchange interaction $J_\perp \simeq 20$~$\mu$eV, obtained at $T\gtrsim 17$~K is comparable to the energy scale $g\mu_B H_{\rm cr}$, corresponding to the ``critical'' field $H_{\rm cr}=0.2$~T of the cusp of the dependence $T_{\rm SP}(H)$ in the phase diagram of Fig. \ref{fig:Phase_map}.
%obtained in Sect. \ref{Sect:SRO}, which is necessary for the onset of SP phase at low temp. 
Therefore, we can associate the obtained cusp with the onset of  interlayer coherence.
%, in contrast to the interpretation given previously in Ref. \cite{kesarwani2025investigation}, which connected this transition with the onset of two-dimensional order. 
%In this respect the obtained behavior $T_{\rm SP}(H)$ as a function of the external magnetic field can be explained as follows. 
Indeed, for $h=g\mu_BH<\alpha J_\perp$ ($\alpha\sim 1$) the magnetic field induces only a small change in the magnetic transition temperature, since the magnetic fluctuations are cut by larger interlayer exchange. This leads to the standard result $T_{\rm SP}(h)\simeq T_{\rm SP}(0)=T_{\rm BKT}+A/\ln^2(B/J_\perp)$ ($A, B$ are some constants) \cite{Hikami,irkhin1999kosterlitz}. At the same time, at $h>\alpha J_\perp$  the magnetic field serves as a cutoff of the easy plane magnetic fluctuations. Then
\begin{equation}
T_{\rm SP}(h)\simeq T_{\rm BKT}+\frac{A}{\ln^2 ({\alpha B}/{h})}\simeq T_{\rm SP}(0)+
T_1 \ln \frac{h}{\alpha J_\perp}
%\frac{2A \ln (h/\alpha J_\perp)}{\ln^3({B}/J_\perp)}}  
\end{equation}
($T_1=2A /{\ln^3({B}/J_\perp)}$) which may explain the linear-like dependence of $T_{\rm SP}(H)$ on $\ln H$ in the phase diagram of Fig.~\ref{fig:Phase_map} {(we note the logarithmic scale of the $H$ axis)}. This also stresses the onset of 3D coherence in small magnetic fields already at the temperature $T_{\rm SP}(0)$, in contrast to the interpretation given previously in Ref. \cite{mcguire2017magnetic}, which connected this transition with the onset of two-dimensional order. 
%We note, that within the Heisenberg model the long-range order has three-dimensional character for arbitrarily small interlayer exchange interaction.

In the low-temperature regime, the interlayer exchange interaction can be extracted from the experimentally measured field $H_E$. The respective exchange interaction is negative, consistent with the 3D AFM ground state. However, as the temperature approaches the transition to the spin-polarized phase near 17 K, the AFM exchange is systematically suppressed. This observation necessitates the existence of a sign-change in the interlayer exchange as a function of temperature.

The change of the sign of interlayer exchange may be closely related to the presence of stacking defects \cite{schneeloch2024role} (we also note the observation of the coexistence of rhombohedral and monoclinic phases down to very low temperatures \cite{kesarwani2025investigation}). It was shown that these defects may change the magnetic behavior in the region 14-17 K, shifting the balance towards ferro- or antiferromagnetic phases. In that respect, we would also like to mention the observation of FM-AFM hysteresis noted in Ref. \cite{Bykovetz2019} and the sensitivity of the magnetic phase to external pressure \cite{Ahmad2020}. These observations show that the balance between ferro- and antiferromagnetic phases is rather fragile and influenced by external conditions, as well as the lattice degrees of freedom. In conjunction with the magnetoelastic coupling, one can expect a switch from ferromagnetic correlations at high temperatures to an antiferromagnetic structure at low temperatures, in agreement with the experimental data. We also note that the peak of the specific heat at $T=T_C\simeq 17 K$ is rather broad \cite{mcguire2017magnetic}, while the uniform susceptibility shows a clear divergence, which may also be considered as additional evidence that the lattice degrees of freedom are involved in the vicinity of this transition. The presence of stacking defects can also naturally explain the paramagnetic nature of the SP phase, with the magnetization vanishing in small fields \cite{kesarwani2025investigation} as a consequence of magnetic disorder in a continuous symmetry system \cite{ImryMa}, {as well as the obtained small critical exponent of correlation length in our EPR study}.

%This crossover likely originates from the delicate balance between FM superexchange mediated by the ligands and AFM direct exchange between chromium ions. \cite{besbes2020microscopic}. 
%The suppression of AFM coupling near the transition allows the strong intralayer ferromagnetic correlations to dominate, facilitating the emergence of the 2D-ordered SP phase before 3D coherence is lost \cite{mcguire2017magnetic}.

\section{Discussion}

We have studied the properties of local magnetic moments and their exchange interactions theoretically above the magnetic transition temperature. Our theoretical study was complemented by measurements of magnetic properties both above and below the magnetic transition temperature. Due to the presence of wide energy gap  (approximately 3.31 eV  \cite{ermolaev2025giant}) the local magnetic moments have long room-temperature lifetime $\tau$ of 130--300 ps, which further increases with decrease of temperature. The in-plane exchange interactions are found rather weakly dependent on temperature.

A central finding of this work is the persistence of robust short-range order well into the paramagnetic phase. Our DFT+DMFT calculations predict a formal divergence of the correlation length $\xi$ at approximately $100$ K, marking the onset of localized FM cluster formation. This theoretical prediction is corroborated by the giant $22\text{ cm}^{-1}$ shift of the $E_g^2$ Raman mode, which begins to deviate from anharmonic behavior at temperatures as high as $80\text{--}100$ K \cite{zhu2025theoretical, kesarwani2025investigation}. This shift is nearly four times larger than that observed in $\rm CrI_3$, indicating an exceptionally strong spin-phonon coupling in $\rm CrCl_3$. The lattice appears to be affected by the formation of FM domains long before they achieve global 3D coherence, implying that the paramagnetic state is actually a fluid of fluctuating 2D-ordered clusters.

{While the intralayer magnetization, extracted from the frequencies of magnetic resonance, becomes small already at $T>30$~K (see Fig. \ref{fig:FMR_temp_dep}a), the width of the magnetic resonance lines (Fig. \ref{fig:FMR_linewidth}) remains non-negligible up to at least $60$~K. Therefore, presence of strong short range magnetic order finds its direct confirmation from the obtained magnetic resonance spectrum. The remaining discrepancy to the estimate of the DFT+DMFT approach can be ascribed to the mean-field nature of that approach, as well as the density-density form of the considered electron interaction.}
%that remains remarkably robust even at 1160 K due to extremely weak damping with a peak half-width $\Gamma$ below 0.2 meV. This giant lifetime is stabilized by the , which eliminates conduction-electron relaxation pathways. The subtle structural transition from the high-temperature C2/m phase to the low-temperature $R\bar{3}$ phase modulates this dynamic response; the stronger nearest-neighbor exchange $J_1$ in the $R\bar{3}$ phase accelerates spin relaxation , slightly shortening the individual moment lifetime while enhancing global spatial stability. These long-lived local fluctuations are essential for the persistence of the electronic gap and robust optoelectronic properties.

%The analysis of our DFT+DMFT calculations and experimental magnetic characterizations reveals that $\rm CrCl_3$ is a prototypical example of a system governed by a delicate balance between competing electronic and structural degrees of freedom. The unique two-step magnetic transition is the most prominent feature of this material. It can be understood through the lens of extreme sensitivity of the exchange pathways to microscopic lattice distortions, a phenomenon we define as a magneto-structurally driven sign reversal of the interlayer coupling.

Unlike its heavier counterparts, $\rm CrBr_3$ and $\rm CrI_3$, which exhibit a single transition into a 3D ferromagnetic state, $\rm CrCl_3$ undergoes a more complex sequence \cite{kim2019evolution}. At $T_{\rm SP} \approx 17$ K, the system establishes robust  ferromagnetic correlations. %In this intermediate regime, the material behaves as a "pseudo-ferromagnet" where layers are internally ordered but mutually uncorrelated due to the vanishingly weak interlayer exchange \cite{klein2019enhancement}. 
This transition is accurately described by the Berezinskii-Kosterlitz-Thouless mechanism, allowed by the material's easy-plane XY anisotropy \cite{bedoyapinto2021intrinsic}. The second stage, occurring at $T_N \approx 14$ K, marks the transition into the antiferromagnetic ground state. 

We argue that the temperature range $T_N<T<T_{\rm SP}$ is likely characterized by the change of the sign of the interlayer coupling, which occurs as a consequence of the interplay of lattice effects (possible change of stacking) and magnetoelastic coupling. This conclusion is confirmed by the observation of FM-AFM hysteresis in Ref. \cite{Bykovetz2019}, the sensitivity of the magnetic phase to external pressure \cite{Ahmad2020}, as well as the variability of reported Néel temperatures across different samples \cite{schneeloch2024role}. We also note that the changing sign of the interlayer exchange interaction was discussed for CrI$_3$, in spite of the magnetic susceptibility anomaly at the structural phase transition \cite{McGuire2015} and the possibility of electrical field switching between ferro- and antiferromagnetic configurations \cite{Jiang2018}.

{The stacking defects likely play important role also at $T>T_{\rm SP}$, as can be traced from the obtained anomalously low critical exponent $\nu$ of the correlation length, as well as the value $b$ in Eq. (\ref{xsiBKT}) for correlation length in XY regime, which may be explained by smoothening the transition by structural disorder effects. These observations also agree with the broadening peaks of specific heat, corresponding to PM-SP transition, observed earlier \cite{mcguire2017magnetic}}.

Our findings, coupled with previous work, underscore that $\rm CrCl_3$ is not a simple magnetic insulator but a highly tunable platform where magnetic phases can be switched via minimal external stimuli, such as strain or electric fields, by engineering the orbital overlap across the van der Waals gap.
%\vspace{0.3cm}

\section{Acknowledgements}
The DMFT calculations, experimental measurements, and analysis of the results are supported by RSF grant 24-12-00186. Development and adaptation of CT-QMC software are supported within the state assignment of the Ministry of
Science and Higher Education of the Russian Federation
for the IMP UB RAS. 
\appendix
\renewcommand\theequation{A\arabic{equation}}
\renewcommand\thefigure{A\arabic{figure}}
\renewcommand\thetable{A\arabic{table}}
\setcounter{equation}{0}
\setcounter{figure}{0}
\setcounter{table}{0}
%\setcounter{page}{1}
%\section*{Supplemental Material \\
%\lowercase{to the paper }``L\lowercase{ocal magnetic moment formation and }K\lowercase{ondo screening in the presence of }H\lowercase{und exchange: the two-band }H\lowercase{ubbard model analysis}"}
%\vspace{-0.4cm}
%\centerline{T. B. Mazitov and A. A. Katanin}
%\vspace{0.5cm} 

%\bibliographystyleApp{unsrt}
%\begin{thebibliography}{2}

%\end{thebibliography}

\end{document}